\documentclass[twocolumn,showpacs,preprintnumbers,amsmath,amssymb]{revtex4}
\usepackage{color}
\usepackage{bm, graphicx, amsmath}
\newtheorem{thm}{Theorem}

\begin{document}

\title{Finding structural anomalies in star graphs using quantum walks}
\author{Seth Cottrell}
\affiliation{Courant Institute of Mathematical Sciences, New York University, 251 Mercer Street, New York, NY 10012}
\author{Mark Hillery}
\affiliation{Department of Physics, Hunter College of the City University of New York, 695 Park Avenue, New York, NY 10065 USA}

\begin{abstract}
We develop a general theory for a quantum-walk search on a star graph.  A star graph has $N$ edges each of which is attached to a central vertex.  A graph $G$ is attached to one of these edges, and we would like to find out to which edge it is attached.  This is done by means of a quantum walk, a quantum version of a random walk.  This walk contains $O(\sqrt{N})$ steps, which represents a speedup over a classical search, which would require $O(N)$ steps.  The overall graph, star plus $G$, is divided into two parts, and we find that for a quantum speedup to occur, the eigenvalues associated with these two parts in the $N\rightarrow\infty$ limit must be the same.  Our theory tells us how the initial state of the walk should be chosen, and how many steps the walk must make in order to find $G$. 
\end{abstract}


\maketitle

A quantum walk is a quantum version of a random walk \cite{reitzner}.  In a quantum walk, a particle moves on a general structure, a graph, which is a collection of vertices and edges connecting them, and its motion is governed by amplitudes, whereas in a classical random walk it would be governed by probabilities.  Quantum walks have proven useful in finding new quantum algorithms.  They have also been realized experimentally in a number of different systems \cite{PeLaPoSoMoSi08} - \cite{schreiber}.

One type of task a quantum walk can perform with a speedup over what can be done classically is a search \cite{shenvi} - \cite{lee}.  Typically one is trying to find a distinguished vertex, that is, one of the vertices of the graph is different from the others, and we would like to find out which one it is.  More recently, it has been found that it is possible to find structural anomalies, such as extra edges, or loops, in graphs by using a quantum walk \cite{feldman,hillery1}.  These papers presented a number of examples of this type of search, and in each case there was a graph with high symmetry, and the anomaly broke that symmetry.  In these examples, there were two features that were unexplained.  First, the graph had to be ``tuned'' in order for the search to work.  That is, certain phases that occur when the particle is reflected from a vertex had to be chosen properly.  Second, only certain initial states for the particle resulted in a quantum speedup.  The theory we present here allows us to understand these features.

Another issue that needs to be addressed is whether a quantum walk search can obtain a speedup that is better than quadratic, that is, if the graph has $N$ vertices, can the walk find the structural anomaly in fewer than $O(\sqrt{N})$ steps.  There is a general proof that the Grover search, which does obtain a quadratic speedup is optimal, but this result does not seem to directly apply to finding structural anomalies in graphs by means of quantum walks \cite{bennett,boyer}.  Here we show, at least for the types of graphs we are considering, that a quadratic speedup is the best one can do.

What we now wish to do is present a more general theory of finding anomalies in highly symmetric graphs.  The basic graph we shall use is a star graph, which has a central vertex with edges radiating from it, and we shall denote this vertex by $0$.  Each of those edges but one is connected to a single vertex (a different vertex for each edge), but one of them is connected to another graph, which we shall call $G$ (see Figure 1).  We shall assume that there are $N$ edges attached to the central vertex with the outer vertices being labelled $1$ through $N$.  We shall also assume, during our analysis, that $G$ is attached to vertex $1$.  In general, we do not know to which vertex $G$ is attached, and our object is to find to which vertex it is, in fact, attached.

Let us now describe our walk in more detail.  We will use a version of a quantum walk known as the scattering walk \cite{hillery2}.  In this walk, the particle sits on the edges, and each edge has two states.  If the edge connects vertices $j$ and $k$, then the state $|j,k\rangle$ corresponds to the particle being on the edge and going from $j$ to $k$, while the state $|k,j\rangle$ corresponds to the particle being on the edge and going from $k$ to $j$.  These states, for all of the edges in the graph, are taken to be orthonormal, and the Hilbert space of the states of the walking particle is just the linear span of these states.  Now that we have a Hilbert space, we need a unitary operator that will advance the walk one step.  In general, each vertex has a unitary operator that maps the states entering the vertex to those leaving it.  The overall unitary that advances the walk one step consists of the joint action of all of these local unitaries.  In the specific case of the star graph with an anomaly, the vertices $2$ through $N$ reflect the particle with a phase, $U|0,j\rangle = e^{i\phi}|j,0\rangle$.  The central vertex behaves as
\begin{equation}
\label{grover-vertex}
U|j,0\rangle = -r|0,j\rangle + t\sum_{k=1,k\neq j}^{N} |0,k\rangle ,
\end{equation}
where $t=2/N$ and $r=1-t$.

We now think of dividing the graph into two pieces, a ``left side'' and a ``right side'' (see Figure 1).  The left side consists of the vertices $2$ through $N$, the edges connected to them, and the central vertex.  The right side consists of $G$, the edge between $0$ and $1$, and the central vertex.  Note that in the $N\rightarrow \infty$ limit, the left and right sides do not communicate.  In this limit, the operator $U$ goes to an operator $U_{0}$, which acts independently on the right and left sides.  Defining
\begin{eqnarray}
|in\rangle = \frac{1}{\sqrt{N-1}}\sum_{j=2}^{N} |j,0\rangle \nonumber \\
|out\rangle = \frac{1}{\sqrt{N-1}}\sum_{j=2}^{N} |0,j\rangle ,
\end{eqnarray}
we have that $U_{0}|in\rangle = |out\rangle$ and $U_{0}|out\rangle = e^{i\phi}|in\rangle$.  This implies that the vectors $(|out\rangle \pm e^{i\phi /2}|in\rangle )/\sqrt{2}$ are eigenstates of $U_{0}$ with eigenvalues $\pm e^{i\phi /2}$, respectively.  We will refer to eigenstates of $U_{0}$ with support only on the left side of the graph as left eigenstates and their eigenvalues as left eigenvalues.  There are also eigenvectors of $U_{0}$ on the right side of the graph.  We will refer to these eigenvectors as right eigenvectors and their eigenvalues as right eigenvalues.  We will be interested in the situation in which one of the right eigenvalues is equal to either $\pm e^{i\phi /2}$.  We will then do perturbation theory in $\epsilon =1/N$ to see what happens in the finite $N$ case.  Note that if we know the right-side eigenvalues, we can always choose $\phi$ so that one of the left-side eigenvalues matches one of the right-side eigenvalues.

In this communication we will present our results without proof, but we will present short arguments to indicate how the results follow.  A longer version of the paper will provide a more detailed discussion and the proofs of the theorems presented here.

\begin{figure}[h]
\centering
\includegraphics[scale=.5]{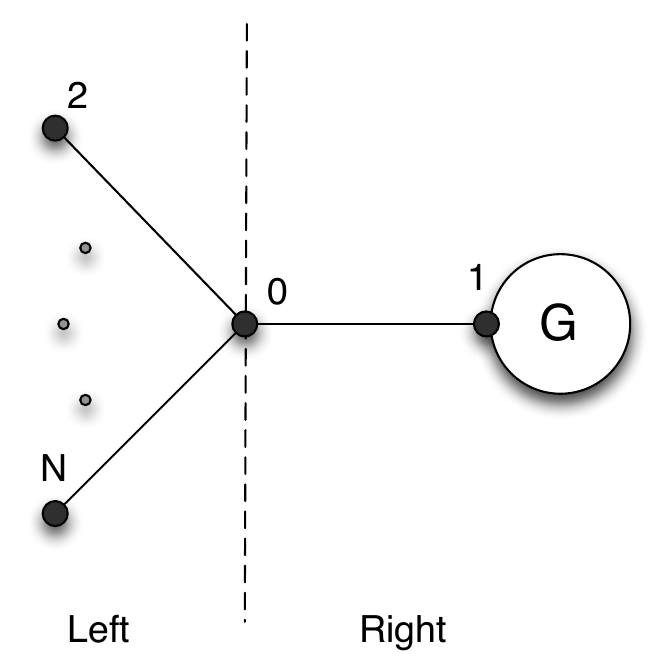}
\caption{A star graph with a second graph, G, attached to one of its external vertices.}
\label{star}
\end{figure}

Let us denote $U$ for a general value of $N$ as $U(\epsilon )$.  With this notation, we have that $U(0) =U_{0}$.  We are interested in the eigenvalues and eigenvectors of $U(\epsilon )$ and, in particular, their dependence on $\epsilon$.  For the eigenvalues we have
\begin{thm}
The eigenvalues, $\lambda$, of the matrix for a quantum walk, $U$, with a characteristic polynomial that is a polynomial in both $\lambda$ and $\epsilon$, can only take the form of $\lambda(\epsilon) = \sum_{j=0}^\infty A_j\epsilon^j$ or $\lambda (\epsilon) = \sum_{j=0}^\infty (-1)^{kj}A_j\left(\sqrt{\epsilon}\right)^j$, where $k=0,1$.
\end{thm}

In order to show where this result comes from we first note that it can be shown that the characteristic polynomial for $U(\epsilon )$ is of the form $C(\lambda ,\epsilon ) = C_{0}(\lambda )+ \epsilon f(\lambda )$.  Suppose that $\lambda_{0}$ is a zero of order $s$ of $C_{0}(\lambda )$ so that $C_{0}(\lambda_{0}+\delta )= a\delta^{s} +\ldots$ and assume that $f(\lambda_{0}+\delta )=b\delta^{q} + \ldots$, where the dots represent higher order terms in $\delta$.  Keeping only lowest order terms, the equation $C(\lambda_{0}+\delta )=0$ gives us $a\delta^{s} + \epsilon b\delta^{q}=0$.  If $q\geq s$, then the solution is just $\delta =0$, and the $s$ eigenvalues are unaffected by the perturbation and remain equal to $\lambda_{0}$.  For $q<s$, we have $\delta = (-\epsilon b/a)^{1/(s-q)}$.  This equation will have $s-q$ nonzero solutions, where the solutions differ by factors of $\exp [2\pi i/(s-q)]$.  However, since $U(\epsilon)$ is unitary for $0\leq \epsilon \leq 1/2$ ($\epsilon = 1/2$ corresponding to $N=2$), the zeroes of $C(\lambda ,\epsilon )$ must lie on the unit circle in the complex plane.  This can only happen if $s-q\leq 2$.  This is because $\lambda_{0}$ is on the unit circle, and if some of the roots split for $\epsilon >0$, then they must remain on the circle, which means that there are only two directions in which they can go.

We can push this a little further.  When $s=q$, we have seen that the $s$ eigenvalues equal to $\lambda_{0}$ when $\epsilon =0$ do not change as $\epsilon$ is increased.  When $s-q=1$, one of the eigenvalues will change and when $s-q=2$, two of them will.  We call the two eigenvalues that result when $\epsilon \neq 0$ in this last case paired eigenvalues.  We can express the paired eigenvalues as $\lambda_\pm \equiv \lambda_0 e^{\pm i c\sqrt{\epsilon}} + O(\epsilon)$, and define $|V^+\rangle$ and $|V^-\rangle$ as the corresponding paired eigenvectors.  As we shall see, the fact that the eigenvalues have the behavior as functions of $\epsilon$ given in the theorem implies that we can obtain at most a quadratic speedup in a search.  The results of the discussion in this and the previous paragraphs lead us to the following theorem.

\begin{thm}[the three-case theorem]
If $\lambda_0$ is a root of $C_0(\lambda)$ with multiplicity $s$, then only one of the following cases applies to the "$\lambda_0$-family" of roots of $C(\lambda,\epsilon)$, $\{\lambda_1(\epsilon), \cdots, \lambda_s(\epsilon)\}$, where $\lambda_k(0)=\lambda_0$, $\forall k$.

i) $\lambda_k(\epsilon)=\lambda_0$, $\forall k$.  That is, all of the roots are constant.

ii) $\lambda_k(\epsilon)=\lambda_0$, for all but one value of $k$.  This root takes the form $\lambda_k(\epsilon)=\lambda_0e^{ib\epsilon}+O(\epsilon^2)$, for some constant $b$.

iii) $\lambda_k(\epsilon)=\lambda_0$, for all but two values of $k$.  These two are paired and take the form $\lambda_\pm = \lambda_0e^{\pm ic\sqrt{\epsilon}}+O(\epsilon)$, for some constant $c$.

\end{thm}

We are interested in case $iii)$, and we would like to know when it will occur.  This is, in fact, the situation that will lead to a search with a quadratic speedup.  The essential idea is that we will get paired eigenvalues when the perturbation, that is $U_{1}=U(\epsilon )-U_0$, couples left and right eigenstates of $U_0$, which we denote by $|L_0\rangle$ and $|R_0\rangle$, respectively, with the same eigenvalue, $\lambda_0$.  The only coupling of these eigenstates occurs through the vertex $0$, the ``hub'' of the graph.  That means that in order to be coupled by the perturbation, $|L_0\rangle$ and $|R_0\rangle$ must have support on edges connected to the hub.  Keeping this in mind leads us to our first result.
\begin{thm}
If $e^{i\phi}+\lambda_{0}^{2}\neq 0$, a right-side eigenvector is constant, and has a constant eigenvalue, if and only if it is not in contact with the hub vertex.
\end{thm}
This is proved in the supplemental material.  We should note that we will assume that $e^{i\phi}+\lambda_{0}^{2}\neq 0$ throughout the rest of this paper.  The eigenvectors, which have support only in $G$, we call bound eigenvectors.  They are eigenvectors of both $U$ and $U_0$.  Eigenvectors of $U_0$ in the space orthogonal to the bound eigenvectors we shall call active eigenvectors.  For a given $\lambda_0$-family, there can be at most two active eigenvectors.  If the $\lambda_{0}$-family contains only left-side or only right-side eigenvectors, then, by the three case theorem, there will be one active eigenvector.  If the $\lambda_{0}$-family contains both left-side and right-side vectors, then pairing is possible.  This leads us to
\begin{thm}[The fundamental pairing theorem]
The $\lambda_0$-eigenspace is in contact with both the left and right sides of the hub vertex if and only if there exists paired vectors $|V^{\pm}\rangle$ with eigenvalues of the form $\lambda_0e^{\pm ic\sqrt{\epsilon}}+O(\epsilon)$, respectively.
\end{thm}
One direction of this theorem is proved in the supplemental material.  Therefore, if the left and right sides share an eigenvalue, $\lambda_0$, of $U_{0}$ and both have an active $\lambda_0$ eigenvector, then there is a pairing.  Another very useful result is
\begin{thm}
Paired eigenvectors are always evenly divided across the hub vertex.  That is, if $P$ is a projection onto either the left or right side and $|V^\pm\rangle$ is a paired eigenvector, then $|\langle V_0^\pm|P|V_0^\pm\rangle| = \frac{1}{2}$.
\end{thm}

One way of seeing how the last two results follow, though not in a fully rigorous way, is to use degenerate perturbation theory.  If the active eigenvectors at $\epsilon = 0$ on the right and left sides are $|R_{0}\rangle$ and $|L_{0}\rangle$, respectively, define the projection operator $Q=|R_{0}\rangle\langle R_{0}| + |L_{0}\rangle\langle L_{0}|$.  We can then define a new unperturbed operator to be $U^{\prime}_{0}= U_{0}+QU_{1}Q$ and the new perturbation to be $U_{1}^{\prime} = U_{1} - QU_{1}Q$.  What we then need to do is diagonalize $U_{0}^{\prime}$.  Looking at the case where $\lambda_{0}=e^{i\phi/2}$, we can express $|L_{0}\rangle$ as 
\begin{equation}
|L_{0}\rangle = \frac{1}{\sqrt{2}} ( |out\rangle + e^{i\phi /2}|in\rangle ) ,
\end{equation}
and the right eigenstate, $|R_{0}\rangle$, as
\begin{equation}
|R_{0}\rangle = \delta |0,1\rangle + \gamma |1,0\rangle + |G\rangle ,
\end{equation}
where $|G\rangle$ is the part of the eigenstate with support in $G$.  The equation $U_{0}|R_{0}\rangle = \lambda_{0}|R_{0}\rangle$ implies that $-\gamma = \lambda_{0}\delta$.  Making use of the relations
\begin{eqnarray}
U_{1}|in\rangle & = & -2\epsilon |out\rangle + 2\sqrt{\epsilon - \epsilon^{2}}|0,1\rangle \nonumber \\
U_{1} |1,0\rangle & = & 2\epsilon |0,1\rangle + 2\sqrt{\epsilon - \epsilon^{2}}|out\rangle ,
\end{eqnarray}
and $U_{1}|out\rangle = U_{1}|0,1\rangle = U_{1}|G\rangle = 0$, we find that in the $|L_{0}\rangle ,\, |R_{0}\rangle$ basis to lowest order in $\epsilon$
\begin{equation}
QU_{1}Q= \lambda_{0}\sqrt{2\epsilon}\left( \begin{array}{cc} 0 & \delta^{\ast} \\ -\delta & 0\end{array} \right) .
\end{equation}
Note that if $\delta =0$, i.e.\ the right eigenstate is not in contact with the hub, the above perturbation vanishes, indicating that there are not paired eigenvectors, a result consistent with Theorem 4. Assuming $\delta\neq 0$ and diagonalizing $U_{0}^{\prime}$, we find that the eigenvectors are
\begin{equation}
|V_{0}^{\pm}\rangle = \frac{1}{\sqrt{2}} \left( \mp \frac{i|\delta |}{\delta}|R_{0}\rangle + |L_{0}\rangle \right) ,
\end{equation}
with corresponding eigenvalues $\lambda_{0} \pm i\sqrt{2} \lambda_{0} |\delta | \sqrt{\epsilon} + O(\epsilon ) = \lambda_{0}e^{\pm ic\sqrt{\epsilon}}+O(\epsilon)$, with $c= \sqrt{2} |\delta |$.  Note that this result and that of the previous equation are the same as the those stated in Theorems 4 and 5.

Since we don't know where the flawed/marked edge is, the initial state is assumed to be composed of a uniform superposition of the form 
\begin{equation}
|\psi_{init}\rangle = \frac{\alpha}{\sqrt{N}}\sum_{j=1}^N|0,j\rangle + \frac{\beta}{\sqrt{N}}\sum_{j=1}^N|j,0\rangle .  
\end{equation}
This initial state is almost entirely concentrated on the left side.  We can choose $\alpha$ and $\beta$ so that the initial state is equal to $|L_0\rangle$ up to $O(\sqrt{\epsilon})$.  In particular, if $\lambda_{0}=e^{i\phi /2}$, and we choose $\alpha =1$ and $\beta = e^{i\phi /2}$, then this will be the case.  We then have that
\begin{eqnarray}
U^{m}|\psi_{init}\rangle & = & \frac{1}{\sqrt{2}} U^{m} [|V^{+}_{0}\rangle + |V^{-}_{0}\rangle  + O(\sqrt{\epsilon}) ] \nonumber \\
 & = & \lambda_{0}^{m} ( e^{imc\sqrt{\epsilon}} |V^{+}_{0}\rangle + e^{-imc\sqrt{\epsilon}} |V^{-}_{0}\rangle ) \nonumber \\
 & & + O(\sqrt{\epsilon})
\end{eqnarray}
When $mc\sqrt{\epsilon}=\pi /2$ the above state is, up to terms of $O(\sqrt{\epsilon})$ and a phase, equal to $|R_{0}\rangle$, that is, it is entirely on the right-hand side.  Note that this happens in $O\left(\sqrt{N}\right)$ iterations of $U$.  At that time a measurement is nearly guaranteed to find a state in the right side, which is the location of the marked edge.  In making a measurement to determine the location of the particle, the only part of the right side to which we have access is the edge between $0$ and $1$.  If we had access to $G$, we would know where it is.  Therefore, the probability of successfully determining where the particle is when it is on the right side depends upon the probability that it is on the edge between $0$ and $1$, i.e.\ up to $O(\sqrt{\epsilon})$ on $|\langle 0,1|R_{0}\rangle |^{2} + |\langle 1,0|R_{0}\rangle |^{2}$.

We can quickly determine $\lambda_0$, $|L_0\rangle$, and $|R_0\rangle$ because all of them are determined entirely by $U_0$, which tends to be easy to work with.  In addition to being block-diagonal (the blocks corresponding to the left and right sides), $U_0$ is often sparse, and unlike $U$, it has no dependence on $\epsilon$.

We can find $c$, and thus $m$, using the fact that $c = \lim_{\epsilon\to0} \frac{1}{\sqrt{\epsilon}}|\langle R_0|U|L_0\rangle|$.  In the particular case of a star graph, $c = \sqrt{2}|\langle1,0|R_0\rangle|$.  The number of iterations of $U$ needed to take $|L_0\rangle$ to $|R_0 \rangle$ is $m=\frac{\pi}{2c\sqrt{\epsilon}}$.

\begin{figure}[h]
\centering
\includegraphics[scale=.5]{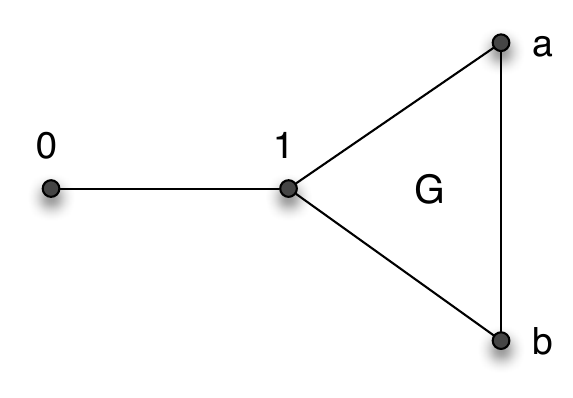}
\caption{A triangle graph attached to an extra edge.  The triangle will play the role of the graph, $G$, in our example.}
\label{triangle}
\end{figure}

Let us conclude with an example.  Suppose that $G$ is a triangle as shown in Fig.\ 2.  The vertices $a$ and $b$ are completely transmitting, e.g.\ $U_{0}|1,a\rangle = |a,b\rangle$, and vertex $1$ behaves as in Eq.\ (\ref{grover-vertex}) with $r=1/3$ and $t=2/3$.  In order to find $\lambda_{0}$ and $|R_{0}\rangle$ we take vertex $0$ to be perfectly reflecting with a phase of $\pi$ so that $U_{0}|1,0\rangle = - |0,1\rangle$.  Defining the vectors 
\begin{eqnarray}
|u_{1}\rangle & = & \frac{1}{\sqrt{2}}(|1,a\rangle + |1,b\rangle ) \nonumber \\
|u_{2}\rangle & = & \frac{1}{\sqrt{2}}(|a,1\rangle + |b,1\rangle ) \nonumber \\
|u_{3}\rangle & = & \frac{1}{\sqrt{2}}(|a,b\rangle + |b,a\rangle ) ,
\end{eqnarray}
we find that $U_{0}|1,0\rangle = -|0,1\rangle$, $U_{0}|u_{1}\rangle = |u_{3}\rangle$, $U_{0}|u_{3}\rangle = |u_{2}\rangle$, and 
\begin{eqnarray}
U_{0}|0,1\rangle & = & -\frac{1}{3}|1,0\rangle + \frac{2\sqrt{2}}{3}|u_{1}\rangle \nonumber \\
U_{0}|u_{2}\rangle & = & \frac{1}{3} |u_{1}\rangle + \frac{2\sqrt{2}}{3}|1,0\rangle .
\end{eqnarray}
Therefore, $S= {\rm span} \{ |0,1\rangle , |1,0\rangle , |u_{1}\rangle , |u_{2}\rangle , |u_{3}\rangle \}$ is invariant under the action of $U_{0}$, and finding the active eigenvectors of $U_{0}$ is reduced to diagonalizing a $5\times 5$ matrix.  Doing so, we find that one of the eigenvalues is $\lambda_{0}=-1$, and the corresponding eigenvector is
\begin{eqnarray}
|R_{0}\rangle & = & \sqrt{\frac{2}{7}}(|0,1\rangle + |1,0\rangle ) - \frac{1}{\sqrt{7}}(|u_{1}\rangle
\nonumber \\
& & + |u_{2}\rangle - |u_{3}\rangle ) .
\end{eqnarray}
We now find that $c=2/\sqrt{7}$, and when the particle is in the state $|R_{0}\rangle$, the probability that it is on the edge between vertices $0$ and $1$ is $4/7$.  Therefore, in order to find the triangle graph attached to a star with $N$ edges, we should set $\phi=2\pi$ (so that $\phi /2=\pi$), start in the state 
\begin{equation}
|\psi_{init}\rangle = \frac{1}{\sqrt{N}}\sum_{j=1}^{N}( |0,j\rangle - |j,0\rangle )  , 
\end{equation}
which is, up to $O(\sqrt{\epsilon})$, equal to the left-side $\lambda_0=-1$ eigenstate,
run the walk for $m= 7\pi \sqrt{N}/8$ steps, and measure the edges of the star to see where the particle is.  With a probability of $4/7$ we will find it on the edge between $0$ and $1$.  If our graph is specified by an adjacency list (a list for each vertex of the other vertices to which it is connected) we can then just check whether the vertex other than the central vertex connected to the edge on which we have found the particle is connected to vertices other than the central vertex.  If the answer is ``yes'', then we are done.  If the answer is ``no'', then we run the walk again.  We will only have to run it $O(1)$ times in order to find to which vertex $G$ is attached. 

Let us now summarize our conclusions.  First, for a quantum walk search whose object is to find a general structural anomaly, $G$, attached to the hub of a star graph, a quadratic speedup is the best that can be done.  Second, for there to be a quadratic speedup, the reflection phases of the external vertices to which $G$ is not attached need to be chosen so that the left and right sides of the graph share a common eigenvalue, i.e.\ there are two eigenstates of $U_{0}$, one completely on the left side and one completely on the right side, that have the same eigenvalue.  Finally, the initial state of the walk must be chosen to be close to the left-side eigenstate.  

The approach outlined here can be pushed in a number of directions.  One can ask how close the eigenvalues of the parts of the graph have to be for a quadratic speedup to take place.  It would be useful to know if anything can be said about properties of $G$ and how much of an active eigenstate lies on the edge between vertices $0$ and $1$ (a lower bound is placed on this in the supplemental material).  Another possibility that can be investigated is when more than one spoke from the star goes into the graph, $G$.  Finally, it may be possible to use this type of an approach to determine whether two graphs are connected; if the graphs have equal eigenvalues, it might be possible for a particle starting in one graph to make it to the other, if they are connected, with some kind of a quantum speedup.  This, however, needs much further work.

\section*{Supplemental material}
\subsection{Proof of Theorem 3}
One direction of the theorem is straightforward.  If the right-side eigenvector is not in contact with the hub, it is localized in $G$, and $U$ and $U_{0}$ will have the same action on the eigenvector.  Therefore, even for $\epsilon\neq 0$, it will be an eigenvector of $U$ with eigenvector $\lambda_{0}$.

The other direction requires more work.  We start by writing the right-hand eigenstate of $U_{0}$ as
\begin{equation}
|V_{0}\rangle = \gamma_{0}|1,0\rangle + \delta_{0}|0,1\rangle + |G_{0}\rangle ,
\end{equation}
where $|G_{0}\rangle$ has support in $G$.  The equation $U_{0}|V_{0}\rangle = \lambda_{0}|V_{0}\rangle$ gives us that $\gamma_{0}=-\delta_{0}\lambda_{0}$ and
\begin{equation}
\label{u0-eq}
U_{0}[ \delta_{0}|0,1\rangle + |G_{0}\rangle ] =  \lambda_{0} (\gamma_{0}|0,1\rangle + |G_{0}\rangle ) .
\end{equation}
We now move to the $\epsilon\neq 0$ case. The statement that $|V(\epsilon )\rangle$ is a right-side eigenvector means that it is an eigenvector of $U$ satisfying $|V(0)\rangle = |V_{0}\rangle$, i.e.\ at $\epsilon = 0$ is is entirely on the right side (this rules out paired eigenvectors, which have components on the left and right sides in the $\epsilon \rightarrow 0$ limit).  We can express $|V\rangle$ (we shall not explicitly note the $\epsilon$ dependence) as
\begin{eqnarray}
|V\rangle & = & \alpha |in\rangle + \beta |out\rangle + \gamma |1,0\rangle \nonumber \\
& & + \delta |0,1\rangle + |G\rangle ,
\end{eqnarray}
where $|G\rangle$ has support in $G$.  We then have that
\begin{eqnarray}
U|V\rangle & = & \alpha [(1-2\epsilon )|out\rangle + 2\sqrt{\epsilon -\epsilon^{2}} |0,1\rangle ] \\
& & + \gamma [(-1+2\epsilon )|0,1\rangle + 2\sqrt{\epsilon - \epsilon^{2}} |out\rangle ] \\
& & + \beta e^{i\phi}|in\rangle + U[\delta |0,1\rangle + |G\rangle ] .
\end{eqnarray}
If we assume that $U|V\rangle = \lambda_{0}|V\rangle$, i.e.\ that the eigenvalue is independent of $\epsilon$, we get three equations relating the four quantities $\alpha$, $\beta$, $\gamma$, and $\delta$, and the equation
\begin{equation}
\label{u-eq}
U[\delta |0,1\rangle + |G\rangle ] = \lambda_{0} (\gamma |1,0\rangle + |G\rangle ) .
\end{equation}
We can now make use of the fact that $U[\delta |0,1\rangle + |G\rangle ]=U_{0}[\delta |0,1\rangle + |G\rangle ] $, since the action of $U$ on $|0,1\rangle$ and $|G\rangle$ is unaffected by the hub vertex.  Therefore, one can replace $U$ by $U_{0}$ on the left-hand side of the above equation, and then take the inner product of Eqs.\ (\ref{u0-eq}) and (\ref{u-eq}).  This results in the relation, $-\delta \lambda_{0}=\gamma$, which is a fourth equation for our four unknown quantities.  Eliminating variables, we find that $\beta$, $\gamma$, and $\delta$ are proportional to $\alpha$, and
\begin{equation}
\alpha (\lambda_{0}^{2}+e^{i\phi}) = 0 .
\end{equation}
This implies that if $\lambda_{0}^{2}+e^{i\phi} \neq 0$, then $\alpha =\beta =\gamma =\delta =0$.  Therefore, if the eigenvalue is independent of $\epsilon$, and $\lambda_{0}^{2}+e^{i\phi} \neq 0$, then the eigenvector is not in contact with the hub.

\subsection{Proof of one direction of Theorem 4}
We will prove the following:
\begin{thm}
Paired eigenvectors straddle the hub, in the sense that if $|V\rangle$ is a paired eigenvector, then the projection of $|V_0\rangle$ onto either the left or right side is non-zero, where $|V\rangle = |V_{0}\rangle + \sqrt{\epsilon}|V_{1}\rangle + O(\epsilon)$ and $U_{0}|V_{0}\rangle = \lambda_{0}|V_{0}\rangle$ and $\|V\| = \| V_{0}\| = 1$.
\end{thm}
Assuming that $\langle V|V\rangle=1$ and $\lambda = \lambda_0 e^{\pm ic\sqrt{\epsilon}}+O(\epsilon)$, we find immediately that, $\langle V|U|V\rangle = \lambda$ implies that
\begin{eqnarray}
\langle V_1|U_0|V_0\rangle + \langle V_0|U_0|V_1\rangle + \langle V_0|\tilde{U}_1|V_0\rangle & = & \pm ic\lambda_0 \nonumber \\
\lambda_0\left[\langle V_1|V_0\rangle + \langle V_0|V_1\rangle\right] + \langle V_0|\tilde{U}_{1}|V_0\rangle & = & \pm ic\lambda_0 ,
\end{eqnarray}
where $U=U_{0} + \sqrt{\epsilon}\tilde{U}_{1} +O(\epsilon )$.
Now $\langle V|V\rangle = 1$ and $\langle V_{0}|V_{0}\rangle = 1$ imply that $\langle V_{0}|V_{1}\rangle = 0$, so we have that
\begin{equation}
\langle V_0|\tilde{U}_1 |V_0\rangle = \pm ic\lambda_0 .
\end{equation}
But $\tilde{U}_1$ is only involved in transmission across the hub, so when $c\ne 0$ we find that $|V_0\rangle$ must have some component on both sides.  This requires that the $\lambda_0$-eigenspace of $U_0$ appears on both sides of the hub as well.
$\square$

\subsection{Lower bound on $c$}
One of the factors determining how many steps a search will take is the size of $c$, which for a star graph is $\sqrt{2}|\langle 1,0|R_{0}\rangle |$, and $|R_{0}\rangle$ is a right-side active eigenvector.  The number of steps is inversely proportional to $c$, so if $c$ is small, the search will require many steps.  We can put a lower bound on $c$ in terms of the number of edges in the graph $G$.  Let $d$ be the number of right-side active eigenvectors, $\{ |R_{n}\rangle\, |\, n=1,2,\ldots d\}$.  We have that
\begin{equation}
1=\sum_{j=1}^{d} |\langle 1,0|R_{n}\rangle |^{2} ,
\end{equation}
because only the active eigenvectors have an overlap with $|1,0\rangle$.  This implies that at least one of the terms in the sum is greater than or equal to $1/d$, and $d$ cannot be any larger than the dimension of the right-side subspace, which is $2E+2$, where $E$ is the number of edges in $G$.  Therefore, there is an active eigenvector for which
\begin{equation}
c \geq \sqrt{\frac{1}{E+1}} .
\end{equation}

\end{document}